\documentclass[aps,prl,twocolumn,showpacs,superscriptaddress]{revtex4}

\usepackage{graphicx}
\usepackage{amsmath}
\usepackage{amssymb}

\input{epsf}
\begin{document}

\title{Dipolar Bose gases: Many-body versus mean-field description}

\author{D.C.E.  Bortolotti}
\affiliation{JILA, NIST and Department of Physics, University of Colorado,
Boulder, CO 80309-0440}
\affiliation{LENS and Dipartimento di Fisica, Universit\'a di Firenze,
 Sesto Fiorentino, Italy}
\author{S. Ronen}
\author{J.L. Bohn}
\affiliation{JILA, NIST and Department of Physics, University of Colorado,
Boulder, CO 80309-0440}
\author{D. Blume}
\affiliation{Department of Physics and Astronomy, 
Washington State University,
  Pullman, Washington 99164-2814}

\date{\today}

\begin{abstract}
We characterize zero-temperature
dipolar Bose gases 
under external spherical 
confinement  as a function of the
dipole strength using the essentially exact many-body 
diffusion Monte
Carlo (DMC) technique. 
We show that the DMC energies are reproduced accurately
within a mean-field framework if the variation of the $s$-wave scattering
length with the dipole strength is accounted for
properly.
Our calculations suggest 
stability diagrams and collapse mechanisms of dipolar Bose gases 
that differ significantly from those previously proposed in the literature.
\end{abstract}


\maketitle

The recently achieved Bose-Einstein condensation of atomic chromium
\cite{Griesmaier05PRL,Griesmaier05condmat}
has added two new twists to the study of ultracold matter.
First, Cr condensates
realize the first spin-three spinor condensate
\cite{Yi04,Santos05}.  Second, they exhibit, due to
Cr's comparatively large magnetic dipole moment, 
observable anisotropic long-range interactions
\cite{Stuhler05}.
These long-range interactions allow
the relative orientation between well separated atoms or 
molecules to be controlled, either by tuning external fields,
or else by adjusting trap anisotropy. An extensive
theoretical literature has predicted novel properties
for  these gases. For example,
Roton-like features have been predicted for trapped gases 
\cite{ODell03,Santos03},
along with unique phases such as checkerboard, Mott insulator,
and supersolid phases
\cite{Giovanazzi02PRL,Goral02,Damski03}.

Rapid experimental
progress in cooling and trapping suggests that
condensation of ground state molecules with large
permanent dipole moments, such as OH 
\cite{Meerakker05,Bochinski04},  RbCs
\cite{Sage}, KRb \cite{Wang04}, NH \cite{Egorov04}, 
may be achieved soon.
These species would represent truly strongly interacting
dipoles, with interparticle interaction strengths up to $\sim 10^3$
times stronger than in chromium.  Indeed, the dipolar interactions
could become the dominant energy scale in such systems, 
driving transitions to strongly correlated states of these gases.

Thus far, dipolar Bose gases at zero temperature have been described 
by the mean-field Gross-Pitaevskii (GP) equation. 
In particular, stability diagrams and excitation spectra 
have been derived within this formalism 
\cite{Santos00,Yi00,Goral00,Mackie01,Yi01,Goral02PRA,Yi02,Lushnikov02}.
Somewhat surprisingly, the validity of the 
GP equation for dipolar gases with strong, anisotropic
long-range interactions has not been assessed 
in detail to date. {\it Per se} it is not clear that a Hartree
wave function, as used in the GP framework, can properly describe 
systems interacting through potentials that fall off as
$\pm 1/r^3$ at large interparticle distances. 
Neutral atom-atom interactions,
e.g., fall off as $-1/r^6$ and mean-field treatments are 
shown to predict the properties of dilute atomic Bose gases 
with high accuracy.  This requires, however, replacing the
true interaction by an appropriate Fermi pseudopotential.
For electronic systems with a repulsive $1/r$ interactions,
a Hartree-Fock formalism is a suitable starting point
for computing the electronic structure of atoms and molecules.
An accurate determination of observables, however, often
requires correlation effects beyond those described by a
Hartree-Fock wave function.

To address this issue for dipolar interactions, 
this Letter reports essentially exact many-body
diffusion Monte Carlo (DMC)
calculations for dipolar
Bose gases interacting through realistic two-body model potentials.
Our DMC results indicate that the GP equation
is adequate to describe the gas, {\it provided} that
the pseudo-potential is parameterized in terms of a ``dipole-normalized'' 
$s$-wave scattering length $a(d)$.
Using this renormalized $a(d)$ instead of the
``bare'' $s$-wave scattering length 
suggests distinctly different collapse behaviors and stability diagrams
than proposed in the literature and has important implications
for the experimental realization of dipolar Bose gases.

Consider the Hamiltonian $H$ 
for $N$ interacting bosonic dipoles with mass $m$,
assumed to be polarized along the $z$-axis,  
under external harmonic confinement,
\begin{eqnarray}
H=
\sum_{j=1}^N \left( \frac{-\hbar^2}{2m} \nabla_j^2 
+ \frac{1}{2} m \omega^2 \vec{r}_j^2 \right)
+ \sum_{j < k}^N V(\vec{r}_{jk}),
\label{eq1}
\end{eqnarray}
where
$\omega$ denotes the trapping frequency,
$\vec{r}_j$
the position vector with respect to the trap center
of the $j$th dipole, and
$\vec{r}_{jk}$
the distance vector, $\vec{r}_{jk}=\vec{r}_j-\vec{r}_k$.
We model the boson-boson potential $V(\vec{r})$ 
by a short-range hardcore
with cutoff radius $b$ and a long-range tail 
with dipole moment $d$,
\begin{eqnarray}
\label{eq2}
V(\vec{r}) = \left\{ \begin{array}{ll}
	d^2\frac{1 - 3 \cos^2 \theta} {r^3} & \mbox{if $r \ge b$}\\
	\infty & \mbox{if $r < b$} \end{array}, \right. 
\end{eqnarray}
where $\theta$ denotes the angle between the vector $\vec{r}$ 
and the laboratory $z$-axis.
The length  $D_*=m d^2 / \hbar^2$, 
at which the characteristic two-body potential and kinetic
energies coincide, is used in the following to
characterize the anisotropic long-range interaction.

The Hamiltonian $H$ in Eq.~(\ref{eq1}) applies, in cgs units,
to bosons with either magnetic or electric dipole moments.
Importantly, the induced dipole moments that drive the
interaction can be tuned in either case.  Electric dipoles
can be polarized simply by immersing them in an electric field,
whereas magnetic dipoles may by tuned by the scheme described in
Ref.~\cite{Giovanazzi02}.  Consequently,
the ratio $D_*/b$, and hence the
relative importance of the dipolar interaction compared to the 
short-range interaction, can be changed essentially at will.
This motivates us to
investigate the zero temperature
equilibrium properties of dipolar Bose gases
over a wide range of $D_*/b$, including 
the short-range dominated regime with $D_*/b \ll 1$ and the
long-range dominated regime with $D_*/b \gg 1$.  Note that,
to date, the effects of dipolar interactions have been observed 
experimentally only for atomic Cr with
$D_*/b \approx 0.2$ (taking $b$ to be the $s$-wave scattering length).

\begin{figure}
\centerline{\epsfxsize=3.3in\epsfbox{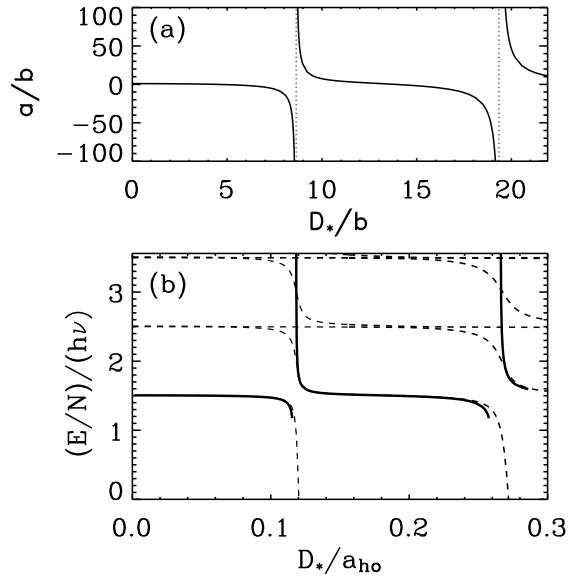}}
\vspace*{-0.6in}
\caption{(a) Solid lines show the $s$-wave scattering length $a$
as a function of the dipole strength $D_*$, both in units of $b$,
 for the two-body potential $V$, Eq.~(\protect\ref{eq2}).
Vertical dotted lines denote those
$D_*/b$ values at which $a$ diverges and a new bound state appears
in the two-body potential.
(b) Dashed lines show $E/N$ for two dipoles under external spherical
confinement calculated for $b=0.0137a_{ho}$ as a function of $D_*/a_{ho}$
obtained by solving the linear Schr\"odinger equation for the Hamiltonian
given by Eq.~(\protect\ref{eq1}).
Solid lines show the corresponding GP energy obtained by
solving Eq.~(\ref{eq4}) for the pseudo-potential 
$V_{\rm eff}$,
Eq.~(\ref{eq3}), using the
dipole-normalized scattering length.
The GP energies are
plotted for each branch of the two-body spectrum.
Note that the $D_*$ values shown in panels (a) and (b)
extend over the same range.
} 
\label{fig1}
\end{figure}

We start our discussion by considering 
two interacting dipoles, i.e., we set $N=2$ in
Eq.~(\ref{eq1}).
After separating off the
center of mass part of $H$, we rewrite the 
Hamiltonian for the relative coordinate in spherical coordinates
and solve the corresponding two-dimensional 
Schr\"odinger equation 
numerically using standard  techniques.
We first determine scattering and bound state solutions in the absence
of an external confining potential, i.e., for $\omega=0$.
Figure~\ref{fig1}(a) shows the 
zero-energy $s$-wave scattering length $a$
as a function of $D_*/b$. We refer to $a$
calculated for $d^2=0$ as the ``bare'' scattering length
and to $a$ calculated for finite $d^2$ as 
the ``dipole-normalized'' scattering length. 
For $D_*=0$, no two-body bound states exist and
$a$ is equal to $b$. 
The scattering length $a$ decreases 
with increasing $D_*$, and 
diverges and changes sign at $D_*/b \approx 8.5$, 
signaling the creation of a two-body bound state.
At $D_*/b \approx 19$, $a$ shows a second divergence
corresponding to a
 second $s$-wave bound state  being pulled in.

Trapped two-body systems could be prepared experimentally by
loading ultracold polar molecules 
into a very deep optical lattice and realizing doubly-occupied lattice
sites.
The two-body energy  for different electric field strengths
can then be measured
spectroscopically.
To determine the energy spectrum of two trapped dipoles 
we fix the short-range two-body length 
$b$, i.e., $b=0.0137 a_{ho}$,
and vary 
$D_*/a_{ho}$, where $a_{ho}$ denotes the oscillator
length, $a_{ho}=\sqrt{\hbar/(m \omega)}$.
Dashed lines in Fig.~\ref{fig1}(b)
show the 
total energy $E/N$ per dipole
as a function
of $D_*/a_{ho}$.
Comparison of Figs.~\ref{fig1}(a) and (b)
reveals that the energetically lowest-lying  state with positive energy
becomes negative at about the same value of $D_*$ 
as that for which the scattering
length $a$ diverges
[the $D_*$ values shown in
panels (a) and (b) of
Fig.~\ref{fig1} extend, although scaled differently, 
over the same range].
Furthermore, 
the trap energies nearly coincide with those for
a non-interacting two-particle gas,
i.e., $E/N=1.5,2.5,\cdots \hbar \omega$, 
at $D_*$ values for which $a=0$. 
The $s$-wave scattering length, which
depends only on the ratio $D_*/b$, thus
determines the gross features
of the energy level spectrum
of two interacting dipoles under external spherical
confinement.
The details of the energy spectrum such as the slope
of the energy levels near $E/N=(n+1/2)\hbar \omega$, however, 
depend 
additionally on the
magnitude of $D_*$ or $b$.

\begin{figure}
\centerline{\epsfxsize=3.3in\epsfbox{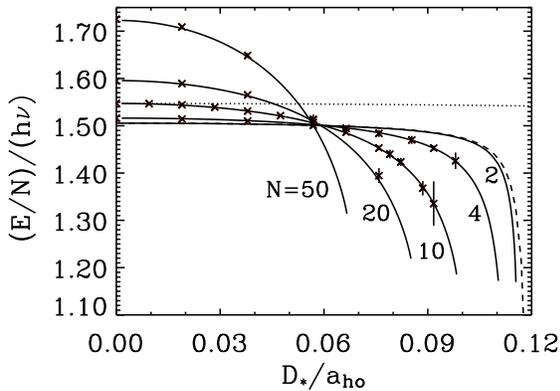}}
\vspace*{-1.6in}
\caption{
Symbols show the energy per particle $E/N$ calculated by the DMC method
for $b=0.0137a_{ho}$
as a function of $D_*/a_{ho}$ for
$N=4$, $10$, $20$ and $50$.
Vertical error bars indicate
statistical uncertainties.
For completeness, a dashed line shows $E/N$, calculated using B-splines,
for $N=2$.
Solid lines show $E/N$ calculated by solving the
non-linear GP equation, Eq.~(\ref{eq4}), with the
dipole-normalized scattering length.
For comparison, a dotted line shows $E/N$
for $N=10$ calculated by solving the
non-linear GP equation, Eq.~(\ref{eq4}), with the
bare scattering length, i.e., the cut-off radius $b$.
} 
\label{fig2}
\end{figure}

For $N>2$, we solve the Schr\"odinger equation using
the DMC technique with importance sampling, which determines the
ground state energy of the
time-independent Schr\"odinger equation 
by propagating an initial ``walker distribution'' 
in imaginary time and projecting out the lowest stationary
eigenstate \cite{Hammond94}.
To efficiently treat 
large systems, a
stochastic realization of the short-time Green's function
propagator is used, which
introduces a
statistical uncertainty
of any DMC expectation value.  Details of the procedure
will be presented elsewhere.
%
Symbols in Fig.~\ref{fig2}
show our DMC energies $E/N$ per dipole 
for $b=0.0137a_{ho}$ (and $d^2$ values for which
$V$ supports no two-body bound states)
as a function of $D_*/a_{ho}$ for
$N=4$, $10$, 
$20$ and $50$.
Statistical uncertainties 
are indicated
by vertical error bars.
For completeness, dashed lines show the $E/N$ data for $N=2$
from Fig.~\ref{fig1}(b).
The energy $E/N$ per dipole decreases with increasing $D_*$.
In particular, $E/N$ becomes
smaller than the ideal gas value of $1.5 \hbar \omega$
for negative $s$-wave scattering lengths ($D_*/a_{\rm ho} $
greater than $\approx 0.06$
in the figure).
Finally, for fixed $D_*/a_{ho}$ the attractive part of the dipolar interaction
leads to a decrease of
$E/N$ with increasing $N$.
We find qualitatively similar behaviors for
dipolar gases confined in elongated cigar-shaped and 
pancake-shaped traps.

Our variational many-body calculations for dipolar gases
show that the region in configuration space where 
the metastable condensate exists is separated by an ``energy barrier''
from the region where bound many-body states exist.
This energy barrier is familiar from variational treatments of
atomic BEC's with attractive interactions 
\cite{PerezGarcia97,Stoof97,Bohn97}.
The existence of this barrier is crucial for
our DMC calculations to converge to the metastable condensate state for
sufficiently large $D_*/a_{ho}$ and not to the cluster-like ground state.
The dipolar gas collapses at the $D_*/a_{ho}$ value
for which the energy barrier vanishes.
Our DMC calculations show that the condensate prior to collapse 
is only slightly elongated, which is consistent with our
finding that the collapse is induced primarily by the negative
value of $a$.

We now assess the validity 
of the GP equation for trapped dipolar Bose gases, which
can be derived by performing a functional variation
of the expectation value
of the Hamiltonian given by Eq.~(\ref{eq1}), calculated with respect
to 
a product wave function $\Psi$,
$\Psi(\vec{r}_1,\cdots,\vec{r}_N)=\prod_{j=1}^N \chi(\vec{r}_j)$.
For this procedure to be meaningful, the 
two-body
interaction potential $V$, Eq.~(\ref{eq2}), has to be replaced by a 
pseudo-potential $V_{\rm eff}$ \cite{Yi00}:
\begin{eqnarray}
\label{eq3}
V_{\rm eff}(\vec{r})= \frac{4 \pi \hbar^2 a(d)}{m} 
\delta(\vec{r}) + 
d^2 \frac{1- 3 \cos ^2 \theta}{{r}^3},
\end{eqnarray}
whose zero-energy T-matrix, calculated in the first Born approximation,
reproduces the full zero-energy
T-matrix of the model potential $V$, Eq.~(\ref{eq2}).
The strength of the contact term of
$V_{\rm eff}$ is not, as might be expected naively,
given by the cutoff radius $b$ but
by the dipole-normalized $s$-wave scattering length $a(d)$.
The 
GP equation for the single particle orbital
$\chi(\vec{r})$ then reads
\begin{eqnarray}
\label{eq4}
\Bigg[ \frac{-\hbar^2}{2m} \nabla^2 +\frac{1}{2} m \omega^2 r^2
+ (N-1)\frac{4 \pi \hbar^2 a(d)}{m} |\chi(\vec{r})|^2 +  \nonumber \\
(N-1) d^2
\int \frac{1-3 \cos ^2 \theta }{|\vec{r}-\vec{r}'|^3} |\chi(\vec{r}')|^2 
d^3 \vec{r}' \Bigg] \chi(\vec{r}) = \epsilon \chi(\vec{r}),
\end{eqnarray}
where $\epsilon$ denotes the chemical potential. 
We solve the non-local
Eq.~(\ref{eq4}) numerically by the steepest
descent method. At each time step, 
the integration over the dipole potential is evaluated
in momentum space with the aid of fast Fourier transforms \cite{Goral00}.
Once the solution to the GP equation is found for a given $N$, $a$
and $d^2$, the 
total energy $E$ can be obtained straightforwardly.

Solid lines in Fig.~\ref{fig2} show the 
GP energies $E/N$ per
dipole
for various $N$ as a function of $D_*/a_{ho}$. 
Figure~\ref{fig2} shows excellent 
agreement between the GP and
DMC energies (symbols) for all $N$ considered. 
To illustrate that this agreement 
depends crucially on the value of the dipole-normalized $s$-wave 
scattering length $a$ in the
contact part of the pseudo-potential $V_{\rm eff}$, Eq.~(\ref{eq3}), 
a dotted line in Fig.~\ref{fig2}
shows the GP energy per dipole for $N=10$
obtained using the cutoff radius $b$
instead of $a$.
Figure~\ref{fig2} indicates that this simple description overestimates
$E/N$ severely when the dipole length $D_*$
becomes comparable to and larger than the short-range length $b$.

\begin{figure}
\centerline{\epsfxsize=2.8in\epsfbox{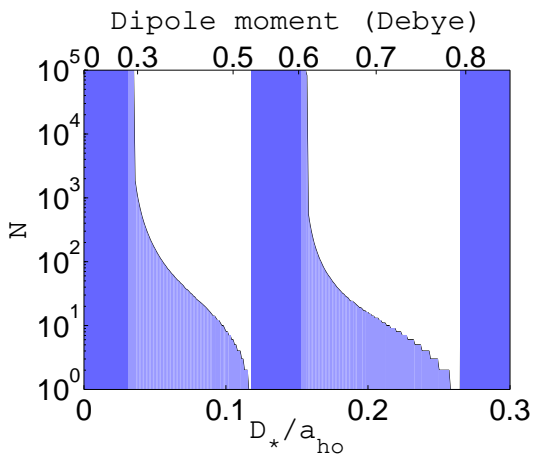}}
\vspace*{-0.0in}
\caption{
Partial stability diagram for a dipolar condensate
The white regions depict parameters for which
the gas is predicted to be mechanically unstable.  Shaded areas 
are regions of stability, and dark shaded areas denote parameters
that are expected to produce stable condensates even in free
space.  The top axis translates the dipole length $D_*$ into
a dipole moment, assuming a trap of frequency $\nu = 1 kHz$
and a molecular mass 20 amu.  } 
\label{fig3}
\end{figure}

The dipole-dependent scattering length has important implications
for the stability of a condensate, as shown in the $N$-vs-dipole
stability  diagram 
in Figure~\ref{fig3}.  For concreteness, we have included an alternative
horizontal axis, representing the dipole moment in Debye, assuming
a trap of frequency $\nu = 1$ kHz, and a molecular mass 20 amu,
typical for light molecules.
The shaded and white areas in Fig.~\ref{fig3} denote parameters 
for which the GP equation does and
does not possess a solution, respectively.  The dark shaded areas
represent where the condensate is expected to be stable for any
number of molecules, even in free space,
as given by the criterion $a(d)>D_*/12\pi$
\cite{Eberlein05}.
Apart from these regions of ``absolute'' stability, the condensate
for a fixed dipole moment will ultimately become unstable as the
number of molecules is increased.  Indeed, for certain values of
dipole where $a(d)$ takes large, negative values 
(say, near $D_*/a_{\rm ho} = 0.27$), a condensate is not supported at all.

Alternatively, for fixed $N$, the condensate stability can be probed
as a function of dipole moment.  This is likely a parameter more
amenable to fine-tuning in the laboratory.  In this case, an initially
stable condensate will collapse after the dipole exceeds a certain
value.  There then follows a region of instability, followed
by another region of stability as the dipole is made yet larger
and the scattering length takes positive values.
This alternating pattern of stable and unstable condensates 
continues beyond the two-and-a-half cycles we have shown in Fig.~\ref{fig3}.
This pattern is in contrast to the generally held view of
polar condensate collapse, which would posit a single collapse
when the dipole reaches a large critical value.  Instead, there are
many critical values, generated each time a new bound state is
absorbed into the two-body potential.  Because this collapse is
largely $s$-wave dominated, the gas is much more nearly
isotropic near the collapse point than has previously been reported.

In summary, we have tested, for the first time, the validity
of the GP equation for describing Bose-Einstein condensates
interacting via strong dipolar forces.  We find that the GP equation
works quite well as compared to essentially exact DMC methods,
as long as the dependence of the $s$-wave scattering length on
dipole moment is accounted for.  Doing so, we predict a rich
stability  diagram for such a system, incorporating
alternating regions of stability and instability as the 
dipole moment is varied.

DB acknowledges financial support from the NSF under grant
No. PHY-0331529, SR from an anonymous fund and from the
U.S.-Israel Educational Foundation (Fulbright Program),
DCEB and JLB from the DOE and the Keck Foundation.

\bibliography{lit}
\bibliographystyle{apalike}

\end{document}